\documentclass[12pt,a4paper]{article}
\makeatletter
\def\eqnarray{%
\stepcounter{equation}%
\let\@currentlabel=\theequation
\global\@eqnswtrue
\global\@eqcnt\z@
\tabskip\@centering
\let\\=\@eqncr
$$\halign to \displaywidth\bgroup\@eqnsel\hskip\@centering
$\displaystyle\tabskip\z@{##}$&\global\@eqcnt\@ne
\hfil$\displaystyle{{}##{}}$\hfil
&\global\@eqcnt\tw@$\displaystyle\tabskip\z@{##}$\hfil
\tabskip\@centering&\llap{##}\tabskip\z@\cr}
\makeatother
\newcommand{\ket}[1]{{\vert{#1}\rangle}}
\newcommand{\bra}[1]{{\langle{#1}\vert}}

\begin{document}

\title{\sl Study on Dynamics of N Level System of Atom \\
by Laser Fields}
\author{
  Kazuyuki FUJII\\
  \thanks{E-mail address : fujii@yokohama-cu.ac.jp}
  Department of Mathematical Sciences\\
  Yokohama City University\\
  Yokohama, 236--0027\\
  Japan
  }
\date{}
\maketitle
%
%
%
%
\begin{abstract}
  This paper is an extension of Fujii et al (quant--ph/0307066) and in this 
  one we again treat a model of atom with n energy levels interacting with 
  n(n-1)/2 external laser fields, which is a natural extension of usual two 
  level system. Then the rotating wave approximation (RWA) is assumed from 
  the beginning.
  
  To solve the Schr{\" o}dinger equation we set the {\bf consistency 
  condition} in our terminology and reduce it to a matrix equation with 
  symmetric matrix $Q$ consisting of coupling constants. 
  
  However, to calculate $\exp(-itQ)$ {\bf explicitly} is not easy. 
  In the case of three and four level systems we determine it in a complete 
  manner, so our model in these levels becomes realistic. 
  
  In last, we make a comment on Cavity QED quantum computation based on 
  three energy levels of atoms as a forthcoming target.
\end{abstract}
%


%
%
%
%

\newpage

\section{Introduction}

The purpose of this paper is to develop a useful method applicable to 
Quantum Computation. As an easy introduction to it see for example 
\cite{LPS}, \cite{nine}, \cite{KF1}.

Quantum Computation is in a usual understanding based on qubits which are 
based on two level system (two energy levels or fundamental spins of atoms), 
See \cite{AE}, \cite{MSIII}, \cite{BR} as for general theory of 
two level system. 

In a realistic image of Quantum Computer we need at least one hundred atoms. 
However, we meet a very severe problem called {\bf Decoherence} which may 
destroy a superposition of quantum states in the process of unitary evolution 
of the system. See for example \cite{WHZ} or recent \cite{MFr1} as 
an introduction. 
At the present time it is not easy to control Decoherence.

By the way, an atom has in general infinitely many energy levels, while 
in a qubit method only two energy levels are used, see for example 
\cite{MFr1} as an introduction of two level approximation. 
We should use this possibility to reduce a number of atoms. Since 
it is not realistic to take all energy levels into consideration at the same 
time we use $n$ energy levels from the ground state, which is in general 
called qudit theory. See for example \cite{KF6}, \cite{KF7}, \cite{KF8}, 
\cite{KuF} and \cite{FFK}. 

In the paper \cite{FHKW1} we considered a model of atom with n energy levels 
interacting with n(n-1)/2 external laser fields, which is a natural extension 
of usual two level system. The rotating wave approximation (RWA) is assumed 
from the beginning. 

In the model it is assumed that all coupling constants regarding 
interactions of atom with different laser fields are equal, which is 
nothing but an approximation theory. To construct a realistic model we again 
treat a ``full" model of atom with n energy levels interacting with n(n-1)/2 
laser fields. 

To solve the Schr{\" o}dinger equation we set the {\bf consistency 
condition} in our terminology and reduce it to a matrix equation with 
symmetric real matrix $Q$ consisting of all coupling constants. 
However, it is not easy to calculate $\exp(-itQ)$ {\bf explicitly}. 

Therefore we restrict to special cases. In the case of three and four 
level systems we determine it in a complete manner, so they become 
realistic. See also \cite{FHKW2}.

In last, we make a comment on Cavity QED quantum computation based on 
three energy levels of atoms that is our forthcoming target.

\section{Review on Two Level System}

Let us review a two level system within our necessity and its Rabi 
oscillation. 
Let $\{\sigma_{1}, \sigma_{2}, \sigma_{3}\}$ be famous Pauli matrices 
\begin{equation}
\sigma_{1} = 
\left(
  \begin{array}{cc}
    0 & 1 \\
    1 & 0
  \end{array}
\right), \quad 
\sigma_{2} = 
\left(
  \begin{array}{cc}
    0 & -i \\
    i & 0
  \end{array}
\right), \quad 
\sigma_{3} = 
\left(
  \begin{array}{cc}
    1 & 0 \\
    0 & -1
  \end{array}
\right), 
\end{equation}
and we set 
\[
\sigma_{+}\equiv \frac{1}{2}(\sigma_{1}+i\sigma_{2})=
\left(
  \begin{array}{cc}
    0 & 1 \\
    0 & 0
  \end{array}
\right), \quad 
\sigma_{-}\equiv \frac{1}{2}(\sigma_{1}-i\sigma_{2})=
\left(
  \begin{array}{cc}
    0 & 0 \\
    1 & 0
  \end{array}
\right). 
\]

Let us consider an atom with two energy levels $E_{0}$ and $E_{1}$ 
($E_{1} > E_{0}$) as two level approximation. 
Its Hamiltonian is in the diagonal form given as 
\begin{equation}
H_{0}=
\left(
  \begin{array}{cc}
    E_{0} & 0     \\
    0     & E_{1}
  \end{array}
\right).
\end{equation}
This is rewritten as 
\[
H_{0}=
E_{0}
\left(
  \begin{array}{cc}
    1 & 0 \\
    0 & 1
  \end{array}
\right)+
(E_{1}-E_{0})
\left(
  \begin{array}{cc}
    0 & 0   \\
    0 & 1
  \end{array}
\right)
=
E_{0}{\bf 1}_{2}+\frac{\Delta}{2}\left({\bf 1}_{2}-\sigma_{3}\right),
\]
where $\Delta=E_{1}-E_{0}$ is the energy difference. 
Since we usually take no interest in constant terms, we can set 
\begin{equation}
\label{eq:2-energy-hamiltonian}
H_{0}=\frac{\Delta}{2}\left({\bf 1}_{2}-\sigma_{3}\right). 
\end{equation}

We consider an atom with two energy levels which interacts with external 
(periodic) field with $g\cos(\omega t+\phi)$. 
In the following we set $\hbar=1$ for simplicity. 
The Hamiltonian in the dipole approximation is given by 
\begin{equation}
\label{eq:2-full-hamiltonian}
H=H_{0}+g \cos(\omega t+\phi)\sigma_{1}
=\frac{\Delta}{2}\left({\bf 1}_{2}-\sigma_{3}\right)+
g\cos(\omega t+\phi)\sigma_{1}, 
\end{equation}
where $\omega$ is the frequency of the external field, $g$ the coupling 
constant between the external field and the atom. This model is complicated 
enough to solve, see \cite{MFr2}, \cite{KF5}, \cite{BaWr}, \cite{SGD}. 

In the following we set $\phi=0$ for simplicity and assume the rotating wave 
approximation (which neglects the fast oscillating terms), namely 
\[
\cos(\omega t)
=\frac{1}{2}(\mbox{e}^{i\omega t}+\mbox{e}^{-i\omega t})
=\frac{1}{2}\mbox{e}^{i\omega t}(1+\mbox{e}^{-2i\omega t})
\approx \frac{1}{2}\mbox{e}^{i\omega t},
\]
and 
\[
\cos(\omega t)\sigma_{1}
=
\left(
  \begin{array}{cc}
    0              & \cos(\omega t) \\
    \cos(\omega t) & 0
  \end{array}
\right)
\approx 
\frac{1}{2}
\left(
  \begin{array}{cc}
    0                     & \mbox{e}^{i\omega t} \\
    \mbox{e}^{-i\omega t} & 0
  \end{array}
\right), 
\]
therefore the Hamiltonian is given by 
\begin{equation}
\label{eq:sub-hamiltonian}
H=
\frac{\Delta}{2}\left({\bf 1}_{2}-\sigma_{3}\right)+\frac{g}{2}
\left(\mbox{e}^{i\omega t}\sigma_{+}+\mbox{e}^{-i\omega t}\sigma_{-}\right)
\equiv 
\frac{\Delta}{2}\left({\bf 1}_{2}-\sigma_{3}\right)+
g\left(\mbox{e}^{i\omega t}\sigma_{+}+\mbox{e}^{-i\omega t}\sigma_{-}\right)
\end{equation}
by the redefinition of $g$ ($g/2 \longrightarrow g$). 
It is explicitly 
\begin{equation}
\label{eq:reduction-matrix}
H=
\left(
  \begin{array}{cc}
    0                      & g\mbox{e}^{i\omega t} \\
    g\mbox{e}^{-i\omega t} & \Delta
  \end{array}
\right). 
\end{equation}

We would like to solve the Schr{\" o}dinger equation 
\begin{equation}
\label{eq:schrodinger-equation}
i\frac{d}{dt}\Psi=H\Psi.
\end{equation}
For that purpose let us decompose $H$ in (\ref{eq:reduction-matrix}) into 
\begin{equation}
\label{eq:decomposition}
\left(
  \begin{array}{cc}
    0                      & g\mbox{e}^{i\omega t} \\
    g\mbox{e}^{-i\omega t} & \Delta
  \end{array}
\right)
=
\left(
  \begin{array}{cc}
    1 &                       \\
      & \mbox{e}^{-i\omega t}
  \end{array}
\right)
\left(
  \begin{array}{cc}
    0 & g      \\
    g & \Delta
  \end{array}
\right)
\left(
  \begin{array}{cc}
    1 &                      \\
      & \mbox{e}^{i\omega t}
  \end{array}
\right),
\end{equation}
so if we set 
\begin{equation}
\label{eq:transformation}
\Phi=
\left(
  \begin{array}{cc}
    1 &                      \\
      & \mbox{e}^{i\omega t}
  \end{array}
\right)
\Psi
\quad \Longleftrightarrow \quad 
\Psi=
\left(
  \begin{array}{cc}
    1 &                       \\
      & \mbox{e}^{-i\omega t}
  \end{array}
\right)
\Phi
\end{equation}
then it is not difficult to see 
\begin{equation}
\label{eq:schrodinger-equation-2}
i\frac{d}{dt}\Phi=
\left(
  \begin{array}{cc}
    0 & g             \\
    g & \Delta-\omega
  \end{array}
\right)
\Phi,
\end{equation}
which is easily solved. For simplicity we set the resonance condition 
\begin{equation}
\label{eq:resonance}
\Delta=\omega,
\end{equation}
then the solution of (\ref{eq:schrodinger-equation-2}) is 
\[
\Phi(t)=
\exp
\left\{-igt
\left(
  \begin{array}{cc}
    0 & 1   \\
    1 & 0 
  \end{array}
\right)
\right\}
\Phi(0)=
\left(
  \begin{array}{cc}
    \cos(gt)  & -i\sin(gt)  \\
   -i\sin(gt) & \cos(gt) 
  \end{array}
\right)
\Phi(0).
\]
As a result, the solution of the equation (\ref{eq:schrodinger-equation}) is 
given as 
\begin{equation}
\Psi(t)
=
\left(
  \begin{array}{cc}
    1 &                       \\
      & \mbox{e}^{-i\omega t}
  \end{array}
\right)
\Phi(t)
=
\left(
  \begin{array}{cc}
    1 &                       \\
      & \mbox{e}^{-i\omega t}
  \end{array}
\right)
\left(
  \begin{array}{cc}
    \cos(gt)   & -i\sin(gt)  \\
    -i\sin(gt) & \cos(gt) 
  \end{array}
\right)
\Phi(0)
\end{equation}
by (\ref{eq:transformation}). If we choose 
$
\Phi(0)=\left(1,\ 0\right)^{T}
$
 as an initial condition, then 
\begin{equation}
\Psi(t)=
\left(
  \begin{array}{c}
    \cos(gt)                       \\
  -i\mbox{e}^{-i\omega t}\sin(gt)
  \end{array}
\right).
\end{equation}
This is a well--known model of the Rabi oscillation (or coherent 
oscillation).

\section{General Theory of N Level System}

In general, an atom has an infinitely many energy levels, however it is not 
realistic to consider all of them at the same time. Therefore we take 
only $n$ energy levels from the ground state into consideration 
($E_{n-1} > \cdots > E_{1} > E_{0}$). Then from the lesson in the preceding 
section the energy Hamiltonian can be 
written as 
\begin{equation}
\label{eq:}
H_{0}=
\left(
  \begin{array}{cccccc}
    0&           &           &       &      &              \\
     &\Delta_{1} &           &       &      &              \\
     &           &\Delta_{2} &       &      &              \\
     &           &           & \cdot &      &              \\
     &           &           &       &\cdot &              \\
     &           &           &       &      & \Delta_{n-1} 
  \end{array}
\right),
\end{equation}
where $\Delta_{j}=E_{j}-E_{0}$ for $1 \leq j \leq  n-1$. For this atom 
we consider the interaction with $n(n-1)/2$ independent external laser 
fields corresponding to every energy difference ($E_{j}-E_{i}$ for $j > i$). 
For example see the following figure for the case of $n=4$ :

\begin{center}
 \input{atom-with-4-energy-levels.fig}
\end{center}
\begin{center}
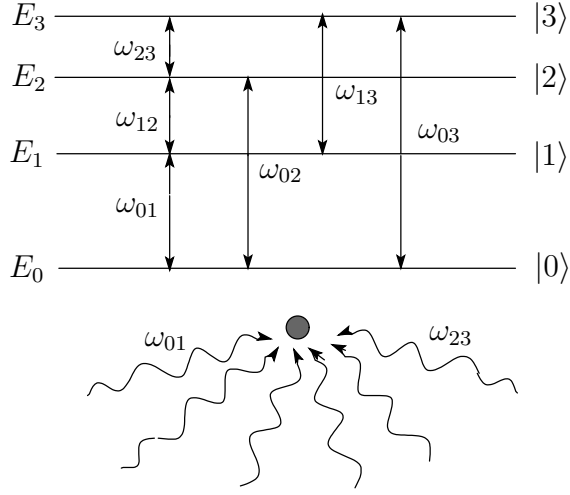
 Fig.1\ Atom with four energy levels and general action by laser fields
\end{center}

\vspace{5mm}
Then the interaction term assuming the RWA from the beginning 
is given as 
\begin{eqnarray}
\label{eq:}
&&V=   \nonumber \\
&&
\left(
  \begin{array}{cccccc}
   0 & g_{01}\mbox{e}^{i\omega_{01}t} & g_{02}\mbox{e}^{i\omega_{02}t} & 
   \cdots & \cdot & g_{0,n-1}\mbox{e}^{i\omega_{0,n-1}t} \\
   g_{01}\mbox{e}^{-i\omega_{01}t} & 0 & g_{12}\mbox{e}^{i\omega_{12}t} & 
   \cdots & \cdot & g_{1,n-1}\mbox{e}^{i\omega_{1,n-1}t} \\
   g_{02}\mbox{e}^{-i\omega_{02}t}& g_{12}\mbox{e}^{-i\omega_{12}t} & 0 & 
   \cdots & \cdot & g_{2,n-1}\mbox{e}^{i\omega_{2,n-1}t} \\
   \cdot& \cdot & \cdot & \cdots & \cdot & \cdot \\
   \cdot& \cdot & \cdot & \cdots & 0 & 
   g_{n-2,n-1}\mbox{e}^{i\omega_{n-2,n-1}t} \\
   g_{0,n-1}\mbox{e}^{-i\omega_{0,n-1}t} & 
   g_{1,n-1}\mbox{e}^{-i\omega_{1,n-1}t} & 
   g_{2,n-1}\mbox{e}^{-i\omega_{2,n-1}t} & \cdots &
   g_{n-2,n-1}\mbox{e}^{-i\omega_{n-2,n-1}t} & 0 
  \end{array}
\right), \nonumber \\
&&
\end{eqnarray}
where $\{g_{\alpha\beta}\}$ are coupling constants, so the Hamiltonian 
is 
\begin{equation}
\label{eq:}
H=H_{0}+V.
\end{equation}
Here we set 
\[
\omega_{1}=\omega_{01},\ \omega_{2}=\omega_{12},\ \cdots, \ 
\omega_{n-2}=\omega_{n-3,n-2},\ \omega_{n-1}=\omega_{n-2,n-1} 
\]
for simplicity. 
We would like to solve the Schr{\" o}dinger equation 
\begin{equation}
\label{eq:schrodinger-equation-general}
i\frac{d}{dt}\Psi=H\Psi.
\end{equation}
Similarly in (\ref{eq:decomposition}) let us decompose $H$ : 
For 
\begin{equation}
\label{eq:U}
U=
\left(
  \begin{array}{ccccccc}
    1&                             & & & & &  \\
     & \mbox{e}^{-i\omega_{1}t}     & & & &  & \\
     &   & \mbox{e}^{-i(\omega_{1}+\omega_{2})t} & & & &  \\
     & & & \cdot  & & & \\
     & & & & \qquad \cdot &  & \\
     & & & & & \mbox{e}^{-i(\omega_{1}+\omega_{2}+\cdots +\omega_{n-2})t} & \\
     & & & & & & \mbox{e}^{-i(\omega_{1}+\omega_{2}+\cdots +\omega_{n-1})t}
  \end{array}
\right)
\end{equation}
it is easy to see 
\begin{eqnarray}
\label{eq:}
&&H_{U}\equiv U^{\dagger}HU=  \nonumber \\
&&
\left(
  \begin{array}{ccccccc}
    0& g_{01} & g_{02}\mbox{e}^{i\epsilon_{02}t} &\cdot & \cdot 
    & g_{0,n-2}\mbox{e}^{i\epsilon_{0,n-2}t} & 
    g_{0,n-1}\mbox{e}^{i\epsilon_{0,n-1}t} \\
    g_{01}& \Delta_{1} & g_{12} & \cdot & \cdot 
    & g_{1,n-2}\mbox{e}^{i\epsilon_{1,n-2}t} & 
    g_{1,n-1}\mbox{e}^{i\epsilon_{1,n-1}t} \\
    g_{02}\mbox{e}^{-i\epsilon_{02}t}& g_{12} & \Delta_{2} & \cdot
    & \cdot & g_{2,n-2}\mbox{e}^{i\epsilon_{2,n-2}t} & 
    g_{2,n-1}\mbox{e}^{i\epsilon_{2,n-1}t} \\
    \cdot & \cdot & \cdot & \cdot & \cdot & \cdot & \cdot \\
    \cdot & \cdot & \cdot & \cdot & \cdot & \cdot & \cdot \\
    g_{0,n-2}\mbox{e}^{-i\epsilon_{0,n-2}t} & 
    g_{1,n-2}\mbox{e}^{-i\epsilon_{1,n-2}t} & 
    g_{2,n-2}\mbox{e}^{-i\epsilon_{2,n-2}t} & \cdot & \cdot & 
    \Delta_{n-2} & g_{n-2,n-1}\mbox{e}^{-i\epsilon_{n-2,n-1}t} \\
    g_{0,n-1}\mbox{e}^{-i\epsilon_{0,n-1}t} & 
    g_{1,n-1}\mbox{e}^{-i\epsilon_{1,n-1}t} & 
    g_{2,n-1}\mbox{e}^{-i\epsilon_{2,n-1}t} & \cdot & \cdot & 
    g_{n-2,n-1}\mbox{e}^{-i\epsilon_{n-2,n-1}t} & \Delta_{n-1}
  \end{array}
\right),  \nonumber \\
&&{} 
\end{eqnarray}
where 
\[
\epsilon_{ij}=\omega_{ij}-(\omega_{i+1}+\omega_{i+2}+\cdots+\omega_{j})
\]
for $j-i\geq 2$. 

By setting 
\[
\tilde{\Psi}=U^{\dagger}\Psi\ \Longleftrightarrow \ \Psi=U\tilde{\Psi}
\]
it is not difficult to see 
\begin{equation}
\label{eq:second-schrodinger-equation}
i\frac{d}{dt}\tilde{\Psi}=\tilde{H}_{U}\tilde{\Psi},
\end{equation}
where 
\begin{eqnarray}
\label{eq:non-diagonal}
&&\tilde{H}_{U}=          \nonumber \\
&&\left(
  \begin{array}{ccccccc}
    0 & g_{01} & g_{02}\mbox{e}^{i\epsilon_{02}t} & \cdot & \cdot 
    & g_{0,n-2}\mbox{e}^{i\epsilon_{0,n-2}t} & 
    g_{0,n-1}\mbox{e}^{i\epsilon_{0,n-1}t}  \\
    g_{01} & \Delta_{1}-\omega_{1} & g_{12} & \cdot & \cdot 
    & g_{1,n-2}\mbox{e}^{i\epsilon_{1,n-2}t} & 
    g_{1,n-1}\mbox{e}^{i\epsilon_{1,n-1}t} \\
    g_{0,2}\mbox{e}^{-i\epsilon_{02}t} & g_{1,2} &
    \Delta_{2}-(\omega_{1}+\omega_{2})& \cdot
    & \cdot & g_{2,n-2}\mbox{e}^{i\epsilon_{2,n-2}t} & 
    g_{2,n-1}\mbox{e}^{i\epsilon_{2,n-1}t} \\
    \cdot& \cdot & \cdot & \cdot & \cdot& \cdot  &\cdot \\
    \cdot& \cdot  & \cdot &\cdot & \cdot & \cdot & \cdot \\
    g_{0,n-2}\mbox{e}^{-i\epsilon_{0,n-2}t} & 
    g_{1,n-2}\mbox{e}^{-i\epsilon_{1,n-2}t} & 
    g_{2,n-2}\mbox{e}^{-i\epsilon_{2,n-2}t} & \cdot & \cdot &
    \Delta_{n-2}-\sum_{l=1}^{n-2}\omega_{l} & 
    g_{n-2,n-1}\mbox{e}^{-i\epsilon_{n-2,n-1}t} \\
    g_{0,n-1}\mbox{e}^{-i\epsilon_{0,n-1}t}& 
    g_{1,n-1}\mbox{e}^{-i\epsilon_{1,n-1}t} & 
    g_{2,n-1}\mbox{e}^{-i\epsilon_{2,n-1}t} & \cdot & \cdot & 
    g_{n-2,n-1}\mbox{e}^{-i\epsilon_{n-2,n-1}t} & 
    \Delta_{n-1}-\sum_{l=1}^{n-1}\omega_{l}
  \end{array}
\right).     \nonumber \\
&&{} 
\end{eqnarray}
At this stage we take the resonance conditions 
\begin{eqnarray}
&&\Delta_{1}=\omega_{1},\ \Delta_{2}=\omega_{1}+\omega_{2}, \ \cdots,\ 
\Delta_{n-2}=\sum_{l=1}^{n-2}\omega_{l},\ 
\Delta_{n-1}=\sum_{l=1}^{n-1}\omega_{l}      \nonumber \\
\Longleftrightarrow \ &&\omega_{j}=E_{j}-E_{j-1}\ (j=1,\ 2,\ \cdots,\ n-1) 
\end{eqnarray}
to make the situation simpler. 

\par \noindent
Moreover, we consider a very special case : namely, 
in (\ref{eq:non-diagonal}) we set
\begin{equation}
\epsilon_{ij}=0 \Longleftrightarrow 
\omega_{ij}=\omega_{i+1}+\omega_{i+2}+\cdots+\omega_{j}=E_{j}-E_{i}
\end{equation}
for all $j-i \geq 2$ (see for example the figure 1 once more). 
We call this a {\bf consistency condition}. 
Then $\tilde{H}_{U}$ becomes a constant symmetric matrix ! 
\begin{equation}
\label{eq:matrix Q}
\tilde{H}_{U}=
\left(
  \begin{array}{ccccccc}
    0 & g_{01} & g_{02} & \cdot & \cdot & g_{0,n-2} & g_{0,n-1} \\
    g_{01} & 0 & g_{12} & \cdot & \cdot & g_{1,n-2} & g_{1,n-1} \\
    g_{02} & g_{12} & 0 & \cdot & \cdot & g_{2,n-2} & g_{2,n-1} \\
    \cdot & \cdot  & \cdot & \cdot & \cdot & \cdot & \cdot  \\
    \cdot & \cdot  & \cdot & \cdot & \cdot & \cdot & \cdot  \\
    g_{0,n-2} & g_{1,n-2} & g_{2,n-2} & \cdot & \cdot & 0 & g_{n-2,n-1} \\
    g_{0,n-1} & g_{1,n-1} & g_{2,n-1} & \cdot & \cdot & g_{n-2,n-1} & 0
  \end{array}
\right)\equiv Q.
\end{equation}

It is easy to solve the equation (\ref{eq:second-schrodinger-equation}) 
with constant $Q$
\begin{equation}
i\frac{d}{dt}\tilde{\Psi}=Q\tilde{\Psi},
\end{equation}
whose solution is formally given by 
\begin{equation}
\label{eq:last-form}
\tilde{\Psi}(t)=\exp(-itQ)\tilde{\Psi}(0). 
\end{equation}
As a result we have a general solution
\begin{equation}
\label{eq:general solution}
\Psi(t)=U\exp(-itQ)\Psi(0). 
\end{equation}

Therefore the problem left is to calculate $\exp(-itQ)$ {\bf explicitly}, 
which is however a very hard task. 
In the following, let us treat the special cases $n=3$ and $n=4$ because 
the formula in \cite{FO} becomes helpful in the cases.

\section{Exact Solution in Three Level System}

In this section we consider the case of $n=3$ and calculate $\exp(-itQ)$ 
in a complete manner.

For simplicity we set $\omega_{01}=\omega_{1}$, $\omega_{02}=\omega_{3}$, 
$\omega_{12}=\omega_{2}$ and $g_{01}=g_{1}$, $g_{02}=g_{3}$, 
$g_{12}=g_{2}$. See the following figure.

\vspace{5mm}
\begin{center}
 \input{atom-with-3-energy-levels.fig}
\end{center}
\vspace{-5mm}
\begin{center}
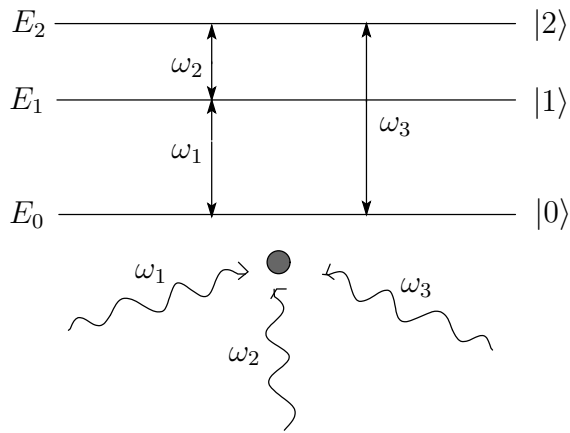
 Fig.2\ Atom with three energy levels and general action by laser fields
\end{center}

Now the matrix $Q$ in (\ref{eq:matrix Q}) is in this case 
\begin{equation}
\label{eq:three-level-matrix Q}
Q=
\left(
  \begin{array}{ccc}
      0   & g_{1} & g_{3} \\
    g_{1} &   0   & g_{2} \\
    g_{3} & g_{2} &  0
  \end{array}
\right),
\end{equation}
so we look for eigenvalues of $Q$. 

From 
\[
0=|\lambda {\bf 1}_{3}-Q|
=
\left|
  \begin{array}{ccc}
    \lambda & -g_{1}  & -g_{3} \\
    -g_{1}  & \lambda & -g_{2} \\
    -g_{3}  & -g_{2}  & \lambda
  \end{array}
\right|
=
\lambda^{3}-\left(g_{1}^{2}+g_{2}^{2}+g_{3}^{2}\right)\lambda-2g_{1}g_{2}g_{3}
\]
the Cardano formula (see for example \cite{KF9}) gives three real solutions 
\begin{equation}
\label{eq:Cardano}
\lambda_{1}=\alpha_{+}+\alpha_{-},\quad 
\lambda_{2}=\sigma^{2}\alpha_{+}+\sigma\alpha_{-},\quad 
\lambda_{3}=\sigma\alpha_{+}+\sigma^{2}\alpha_{-},
\end{equation}
where $\sigma=\mbox{e}^{2\pi i/3}$ and 
\[
\alpha_{\pm}=
\left\{
g_{1}g_{2}g_{3}\pm \sqrt{g_{1}^{2}g_{2}^{2}g_{3}^{2}-
\frac{\left(g_{1}^{2}+g_{2}^{2}+g_{3}^{2}\right)^{3}}{27}}
\right\}^{\frac{1}{3}}.
\]
From \cite{FO} we have finally
\begin{equation}
\exp(-itQ)=f_{0}(t){\bf 1}_{3}+f_{1}(t)Q+f_{2}(t)Q^{2}
\end{equation}
with
\begin{eqnarray}
f_{0}(t)&=&
\frac{\lambda_{2}\lambda_{3}\mbox{e}^{-it\lambda_{1}}}
{(\lambda_{2}-\lambda_{1})(\lambda_{3}-\lambda_{1})}
+
\frac{\lambda_{1}\lambda_{3}\mbox{e}^{-it\lambda_{2}}}
{(\lambda_{1}-\lambda_{2})(\lambda_{3}-\lambda_{2})}
+
\frac{\lambda_{1}\lambda_{2}\mbox{e}^{-it\lambda_{3}}}
{(\lambda_{1}-\lambda_{3})(\lambda_{2}-\lambda_{3})}, \\
f_{1}(t)&=&
-\frac{(\lambda_{2}+\lambda_{3})\mbox{e}^{-it\lambda_{1}}}
{(\lambda_{2}-\lambda_{1})(\lambda_{3}-\lambda_{1})}
-
\frac{(\lambda_{1}+\lambda_{3})\mbox{e}^{-it\lambda_{2}}}
{(\lambda_{1}-\lambda_{2})(\lambda_{3}-\lambda_{2})}
-
\frac{(\lambda_{1}+\lambda_{2})\mbox{e}^{-it\lambda_{3}}}
{(\lambda_{1}-\lambda_{3})(\lambda_{2}-\lambda_{3})}, \\
f_{2}(t)&=&
\frac{\mbox{e}^{-it\lambda_{1}}}
{(\lambda_{2}-\lambda_{1})(\lambda_{3}-\lambda_{1})}
+
\frac{\mbox{e}^{-it\lambda_{2}}}
{(\lambda_{1}-\lambda_{2})(\lambda_{3}-\lambda_{2})}
+
\frac{\mbox{e}^{-it\lambda_{3}}}
{(\lambda_{1}-\lambda_{3})(\lambda_{2}-\lambda_{3})}.
\end{eqnarray}

\section{Exact Solution in Four Level System}

In this section we consider the case of $n=4$ and calculate $\exp(-itQ)$ 
in a complete manner.

For simplicity we also set $\omega_{01}=\omega_{1}$, $\omega_{02}=\omega_{4}$, 
$\omega_{03}=\omega_{6}$, $\omega_{12}=\omega_{2}$, $\omega_{13}=\omega_{5}$, 
$\omega_{23}=\omega_{3}$ 
and $g_{01}=g_{1}$, $g_{02}=g_{4}$, $g_{03}=g_{6}$, $g_{12}=g_{2}$, 
$g_{13}=g_{5}$, $g_{23}=g_{3}$. See the figure 1 once more.

Now the matrix $Q$ in (\ref{eq:matrix Q}) is in this case 
\begin{equation}
\label{eq:four-level-matrix Q}
Q=
\left(
  \begin{array}{cccc}
      0   & g_{1} & g_{4} & g_{6} \\
    g_{1} &   0   & g_{2} & g_{5} \\
    g_{4} & g_{2} &  0    & g_{3} \\
    g_{6} & g_{5} & g_{3} &   0
  \end{array}
\right),
\end{equation}
so we look for the eigenvalues of $Q$ :
\begin{eqnarray*}
0&=&
|\lambda {\bf 1}_{4}-Q|=
\left|
  \begin{array}{cccc}
    \lambda & -g_{1}  & -g_{4}  & -g_{6}  \\
    -g_{1}  & \lambda & -g_{2}  & -g_{5}  \\
    -g_{4}  & -g_{2}  & \lambda & -g_{3}  \\
    -g_{6}  & -g_{5}  & -g_{3}  & \lambda
  \end{array}
\right|   \\
&=&
\lambda^{4}-
\left(g_{1}^{2}+g_{2}^{2}+g_{3}^{2}+g_{4}^{2}+g_{5}^{2}+g_{6}^{2}\right)
\lambda^{2}-
2\left(g_{1}g_{2}g_{4}+g_{1}g_{5}g_{6}+g_{2}g_{3}g_{5}+g_{3}g_{4}g_{6}\right)
\lambda \\
&{}&+\ g_{1}^{2}g_{3}^{2}+g_{2}^{2}g_{6}^{2}+g_{4}^{2}g_{5}^{2}-
2g_{1}g_{2}g_{3}g_{6}-2g_{1}g_{3}g_{4}g_{5}-2g_{2}g_{4}g_{5}g_{6} \\
&\equiv& \lambda^{4}+p\lambda^{2}+q\lambda+r.
\end{eqnarray*}
The Euler formula (see \cite{KF9}) gives four real solutions
\begin{equation}
\label{eq:Euler}
\lambda_{1}=\alpha+\beta+\gamma,\ 
\lambda_{2}=\sigma^{3}\alpha+\sigma^{2}\beta+\sigma\gamma,\ 
\lambda_{3}=\sigma^{2}\alpha+\beta+\sigma^{2}\gamma,\ 
\lambda_{4}=\sigma\alpha+\sigma^{2}\beta+\sigma^{3}\gamma
\end{equation}
with very complicated terms $\alpha$, $\beta$ and $\gamma$ given below 
(corresponding to $\alpha_{+}$, $\alpha_{-}$ in the preceding section)  
and $\sigma=\mbox{e}^{2\pi i/4}=i$. 
For the algebraic equation with three degrees 
\[
64B^{3}+32pB^{2}-4(4r-p^{2})B-q^{2}=0,
\]
we set $\beta=\sqrt{B}$ for the largest real solution $B$ (this can be 
obtained by using the Cardano formula again). $\alpha$ and $\gamma$ are given 
as primitive solutions of equations 
\begin{eqnarray*}
\alpha^{2}&=&\frac{1}{2}
\left\{
-\frac{q}{4\beta}+
\sqrt{\left(\frac{q}{4\beta}\right)^{2}
-\left(B+\frac{p}{2}\right)^{2}}
\right\}, \\
\gamma^{2}&=&\frac{1}{2}
\left\{
-\frac{q}{4\beta}-
\sqrt{\left(\frac{q}{4\beta}\right)^{2}
-\left(B+\frac{p}{2}\right)^{2}}
\right\}. 
\end{eqnarray*}

\par \noindent
From \cite{FO} we have finally
\begin{equation}
\label{eq:last formula}
\exp(-itQ)=f_{0}(t){\bf 1}_{4}+f_{1}(t)Q+f_{2}(t)Q^{2}+f_{3}(t)Q^{3}
\end{equation}
with
\begin{eqnarray}
f_{0}(t)&=&
\frac{\lambda_{2}\lambda_{3}\lambda_{4}\mbox{e}^{-it\lambda_{1}}}
{(\lambda_{2}-\lambda_{1})(\lambda_{3}-\lambda_{1})(\lambda_{4}-\lambda_{1})}
+
\frac{\lambda_{1}\lambda_{3}\lambda_{4}\mbox{e}^{-it\lambda_{2}}}
{(\lambda_{1}-\lambda_{2})(\lambda_{3}-\lambda_{2})(\lambda_{4}-\lambda_{2})}
\nonumber \\
&&+
\frac{\lambda_{1}\lambda_{2}\lambda_{4}\mbox{e}^{-it\lambda_{3}}}
{(\lambda_{1}-\lambda_{3})(\lambda_{2}-\lambda_{3})(\lambda_{4}-\lambda_{3})}
+
\frac{\lambda_{1}\lambda_{2}\lambda_{3}\mbox{e}^{-it\lambda_{4}}}
{(\lambda_{1}-\lambda_{4})(\lambda_{2}-\lambda_{4})(\lambda_{3}-\lambda_{4})}, 
\\
f_{1}(t)&=&
-\frac{(\lambda_{2}\lambda_{3}+\lambda_{2}\lambda_{4}+\lambda_{3}\lambda_{4})
\mbox{e}^{-it\lambda_{1}}}
{(\lambda_{2}-\lambda_{1})(\lambda_{3}-\lambda_{1})(\lambda_{4}-\lambda_{1})}
-
\frac{(\lambda_{1}\lambda_{3}+\lambda_{1}\lambda_{4}+\lambda_{3}\lambda_{4})
\mbox{e}^{-it\lambda_{2}}}
{(\lambda_{1}-\lambda_{2})(\lambda_{3}-\lambda_{2})(\lambda_{4}-\lambda_{2})}
\nonumber \\
&&-
\frac{(\lambda_{1}\lambda_{2}+\lambda_{1}\lambda_{4}+\lambda_{2}\lambda_{4})
\mbox{e}^{-it\lambda_{3}}}
{(\lambda_{1}-\lambda_{3})(\lambda_{2}-\lambda_{3})(\lambda_{4}-\lambda_{3})}
-
\frac{(\lambda_{1}\lambda_{2}+\lambda_{1}\lambda_{3}+\lambda_{2}\lambda_{3})
\mbox{e}^{-it\lambda_{4}}}
{(\lambda_{1}-\lambda_{4})(\lambda_{2}-\lambda_{4})(\lambda_{3}-\lambda_{4})}, 
\\
f_{2}(t)&=&
\frac{(\lambda_{2}+\lambda_{3}+\lambda_{4})\mbox{e}^{-it\lambda_{1}}}
{(\lambda_{2}-\lambda_{1})(\lambda_{3}-\lambda_{1})(\lambda_{4}-\lambda_{1})}
+
\frac{(\lambda_{1}+\lambda_{3}+\lambda_{4})\mbox{e}^{-it\lambda_{2}}}
{(\lambda_{1}-\lambda_{2})(\lambda_{3}-\lambda_{2})(\lambda_{4}-\lambda_{2})}
\nonumber \\
&&+
\frac{(\lambda_{1}+\lambda_{2}+\lambda_{4})\mbox{e}^{-it\lambda_{3}}}
{(\lambda_{1}-\lambda_{3})(\lambda_{2}-\lambda_{3})(\lambda_{4}-\lambda_{3})}
+
\frac{(\lambda_{1}+\lambda_{2}+\lambda_{3})\mbox{e}^{-it\lambda_{3}}}
{(\lambda_{1}-\lambda_{4})(\lambda_{2}-\lambda_{4})(\lambda_{3}-\lambda_{4})},
\\
f_{3}(t)&=&
-\frac{\mbox{e}^{-it\lambda_{1}}}
{(\lambda_{2}-\lambda_{1})(\lambda_{3}-\lambda_{1})(\lambda_{4}-\lambda_{1})}
-
\frac{\mbox{e}^{-it\lambda_{2}}}
{(\lambda_{1}-\lambda_{2})(\lambda_{3}-\lambda_{2})(\lambda_{4}-\lambda_{2})}
\nonumber \\
&&-
\frac{\mbox{e}^{-it\lambda_{3}}}
{(\lambda_{1}-\lambda_{3})(\lambda_{2}-\lambda_{3})(\lambda_{4}-\lambda_{3})}
-
\frac{\mbox{e}^{-it\lambda_{3}}}
{(\lambda_{1}-\lambda_{4})(\lambda_{2}-\lambda_{4})(\lambda_{3}-\lambda_{4})}.
\end{eqnarray}

\section{Approximate Solution of N Level System}

Since we are interested in the whole of $\exp(-itQ)$ in 
(\ref{eq:last-form}) we take an approximation to the equation. Namely, 
we assume that all coupling constants are equal : $g=g_{\alpha\beta}$ for 
$0 \leq \alpha,\ \beta \leq n-1$. Then
\begin{equation}
Q=g
\left(
  \begin{array}{ccccccc}
    0 & 1 & 1 & \cdot  & \cdot & 1  &  1                    \\
    1 & 0 & 1 &  \cdot & \cdot & 1  &  1                    \\
    1 & 1 & 0 & \cdot  & \cdot & 1  &  1                    \\
    \cdot & \cdot  & \cdot & \cdot & \cdot & \cdot & \cdot  \\
    \cdot & \cdot  & \cdot & \cdot & \cdot & \cdot & \cdot  \\
    1 & 1 & 1 & \cdot  & \cdot & 0  &  1                    \\
    1 & 1 & 1 & \cdot  & \cdot & 1  &  0
  \end{array}
\right)\equiv gR
\end{equation}
and it is not difficult to calculate $\exp(-itgR)$ explicitly, \cite{FHKW1}. 

If we define 
\[
\ket{{\bf 1}}=(1,1,\cdots,1,1)^{T},
\]
then it is easy to see 
$
R=\ket{{\bf 1}}\bra{{\bf 1}}-{\bf 1}_{n}
$, 
so 
\[
\exp(-itgR)=e^{igt}\exp(-igt\ket{{\bf 1}}\bra{{\bf 1}}). 
\]
Since $\langle{\bf 1}|{\bf 1}\rangle =n$, 
\[
\left(\ket{{\bf 1}}\bra{{\bf 1}}\right)^{k}
=n^{k-1}\ket{{\bf 1}}\bra{{\bf 1}},
\]
so that 
\begin{eqnarray}
\exp(-igt\ket{{\bf 1}}\bra{{\bf 1}})
&=&{\bf 1}_{n}+
\sum_{k=1}^{\infty}
\frac{(-igt)^{k}}{k!}\left(\ket{{\bf 1}}\bra{{\bf 1}}\right)^{k} \nonumber \\
&=&{\bf 1}_{n}+\sum_{k=1}^{\infty}
\frac{(-igt)^{k}}{k!}n^{k-1} \ket{{\bf 1}}\bra{{\bf 1}} \nonumber \\
&=&
{\bf 1}_{n}+\frac{1}{n}
\left\{\sum_{k=0}^{\infty}\frac{(-ingt)^{k}}{k!}-1\right\}
\ket{{\bf 1}}\bra{{\bf 1}} \nonumber \\
&=&{\bf 1}_{n}+\frac{\exp(-ingt)-1}{n}\ket{{\bf 1}}\bra{{\bf 1}}.
\end{eqnarray}
Therefore we obtain 
\begin{equation}
\exp(-itQ)=\exp(-itgR)=
e^{igt}
\left\{
{\bf 1}_{n}+\frac{\exp(-ingt)-1}{n}\ket{{\bf 1}}\bra{{\bf 1}}
\right\}.
\end{equation}

\section{Discussion}

In this paper we developed a general theory of the $n$ level system of atom 
by assuming the rotating wave approximation and reduce 
the Schr{\" o}dinger equation to the simple matrix equation with matrix 
consisting of coupling constants under the {\bf consistency condition}.

\par \noindent
Moreover, we solved the equation in a perfect manner in the three and four 
level systems.

By the way, to assume the rotating wave approximation is a bit weak in the 
theory, so that we must study the general equation for example in the case of 
three level one like
\begin{equation}
i\frac{d}{dt}
\left(
  \begin{array}{c}
    \Psi_{1}  \\
    \Psi_{2}  \\
    \Psi_{3}
  \end{array}
\right)
=
\left(
  \begin{array}{ccc}
   0 & g_{1}\cos(\omega_{1}t+\phi_{1}) & g_{3}\cos(\omega_{3}t+\phi_{3}) \\
   g_{1}\cos(\omega_{1}t+\phi_{1}) & \Delta_{1} & 
   g_{2}\cos(\omega_{2}t+\phi_{2}) \\
   g_{3}\cos(\omega_{3}t+\phi_{3}) & g_{2}\cos(\omega_{2}t+\phi_{2}) & 
   \Delta_{2} 
  \end{array}
\right)
\left(
  \begin{array}{c}
    \Psi_{1}  \\
    \Psi_{2}  \\
    \Psi_{3}
  \end{array}
\right). 
\end{equation}
However, it is almost impossible to solve this at the present time. 
Let us leave it for young researchers in Quantum Optics or Mathematical 
Physics as a challenging task.

\vspace{3mm}
Here we state our motivation once more. We would like to construct a realistic 
model of quantum computation with three level system. Its candidate is a 
quantum computation based on Cavity QED making use of three energy levels of 
atoms. See the following figure.

\vspace{10mm}
\begin{center}
 \input{cavity-atoms-interaction.fig}
\end{center}
\begin{center}
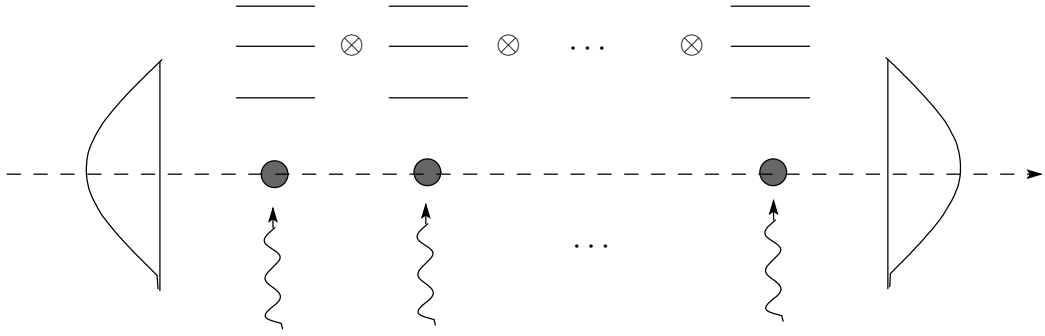
Fig.3\ A general setting for a quantum computation based on Cavity QED
\end{center}

\vspace{5mm}
Let us add a short explanation to this figure. 
The dotted line means a single photon inserted in the cavity and all curves 
mean external laser fields (which are treated as classical ones) subjected to 
ultracold--atoms trapped linearly in it. 
See in detail \cite{FHKW3} and \cite{FHKW4} in which the Cavity QED 
quantum computation with two level system was ``completed".

\par \noindent 
This is our forthcoming target !

\vspace{10mm}
\noindent{\em Acknowledgment.}\\
K. Fujii wishes to thank Akira Asada, Kunio Funahashi and especially 
Shin'ichi Nojiri for their helpful comments and suggestions.

\vspace{10mm}
\begin{center}
\begin{Large}
\noindent{\bfseries Appendix\quad Diagonalization of Q in the Three Level}
\end{Large}
\end{center}
\vspace{5mm}

In the case of three level system let us calculate $\exp(-itQ)$ by use of 
the diagonalization method which is more popular.

In (\ref{eq:Cardano}) we had the three eigenvalues $\{\lambda_{1},\lambda_{2},
\lambda_{3}\}$. Therefore the normalized eigenvectors 
$\{\ket{\lambda_{j}}\ |\ 1\leq j\leq 3\}$ corresponding to these are given as
\begin{equation}
\label{eq:eigenvectors of Q}
\ket{\lambda_{j}}=
\left(
 \begin{array}{c}
  \sqrt{\frac{\lambda_{j}^{2}-g_{2}^{2}}
        {3\lambda_{j}^{2}-\left(g_{1}^{2}+g_{2}^{2}+g_{3}^{2}\right)}} \\
  \sqrt{\frac{\lambda_{j}^{2}-g_{3}^{2}}
        {3\lambda_{j}^{2}-\left(g_{1}^{2}+g_{2}^{2}+g_{3}^{2}\right)}} \\
  \sqrt{\frac{\lambda_{j}^{2}-g_{1}^{2}}
        {3\lambda_{j}^{2}-\left(g_{1}^{2}+g_{2}^{2}+g_{3}^{2}\right)}}
 \end{array}
\right).
\end{equation}
This derivation is given in the latter half. Then the matrix defined by 
\begin{equation}
\label{eq:orthogonal-matrix}
O=\left(\ket{\lambda_{1}},\ket{\lambda_{2}},\ket{\lambda_{3}}\right)
\end{equation}
makes $Q$ diagonal like 
\begin{equation}
\label{eq:diagonal-form of Q}
Q=O
\left(
 \begin{array}{ccc}
   \lambda_{1} & 0 & 0 \\
   0 & \lambda_{2} & 0 \\
   0 & 0 & \lambda_{3}
 \end{array}
\right)
O^{T}.
\end{equation}
Therefore we obtain the desired form
\begin{equation}
\label{eq:last-form}
\exp(-itQ)=
O
\left(
 \begin{array}{ccc}
   \mbox{e}^{-it\lambda_{1}} & 0 & 0 \\
   0 & \mbox{e}^{-it\lambda_{2}} & 0 \\
   0 & 0 & \mbox{e}^{-it\lambda_{3}}
 \end{array}
\right)
O^{T}
\equiv (a_{ij})
\end{equation}
where
\[
a_{ij}=\sum_{k=1}^{3}\mbox{e}^{-it\lambda_{k}}
  \sqrt{\frac{\lambda_{k}^{2}-g_{i+1}^{2}}
       {3\lambda_{k}^{2}-\left(g_{1}^{2}+g_{2}^{2}+g_{3}^{2}\right)}}
  \sqrt{\frac{\lambda_{k}^{2}-g_{j+1}^{2}}
       {3\lambda_{k}^{2}-\left(g_{1}^{2}+g_{2}^{2}+g_{3}^{2}\right)}}\ .
\]

\par \vspace{3mm} \noindent
It is assumed $g_{4}\equiv g_{1}$ in the right hand side.

\par \vspace{3mm}
Let us calculate a special case $g_{3}=0$. From (\ref{eq:Cardano}) 
\[
\lambda_{1}=\sqrt{g_{1}^{2}+g_{2}^{2}},\quad 
\lambda_{2}=0,\quad 
\lambda_{3}=-\sqrt{g_{1}^{2}+g_{2}^{2}}
\]
and from (\ref{eq:eigenvectors of Q}) \footnote{In the calculation we must 
choose signs of the complex roots in (\ref{eq:eigenvectors of Q}) carefully}
\[
\ket{\sqrt{g_{1}^{2}+g_{2}^{2}}}=
\left(
 \begin{array}{c}
  \frac{g_{1}}{\sqrt{2(g_{1}^{2}+g_{2}^{2})}} \\
  \frac{1}{\sqrt{2}} \\
  \frac{g_{2}}{\sqrt{2(g_{1}^{2}+g_{2}^{2})}}
 \end{array}
\right),\quad
\ket{0}=
\left(
 \begin{array}{c}
  \frac{g_{2}}{\sqrt{g_{1}^{2}+g_{2}^{2}}} \\
  0 \\
  \frac{-g_{1}}{\sqrt{g_{1}^{2}+g_{2}^{2}}}
 \end{array}
\right),\quad
\ket{-\sqrt{g_{1}^{2}+g_{2}^{2}}}=
\left(
 \begin{array}{c}
   \frac{g_{1}}{\sqrt{2(g_{1}^{2}+g_{2}^{2})}} \\
  -\frac{1}{\sqrt{2}} \\
   \frac{g_{2}}{\sqrt{2(g_{1}^{2}+g_{2}^{2})}}
 \end{array}
\right).
\]
Therefore from 
\[
O=\left(\ket{\sqrt{g_{1}^{2}+g_{2}^{2}}},\ket{0}, 
\ket{\sqrt{g_{1}^{2}+g_{2}^{2}}}\right)
\]
a straightforward calculation leads us to 
\begin{eqnarray}
\mbox{exp}(-itQ)
&=&
\left(
  \begin{array}{ccc}
  \frac{g_{1}^{2}\mbox{cos}(t\sqrt{g_{1}^{2}+g_{2}^{2}})+g_{2}^{2}}
  {g_{1}^{2}+g_{2}^{2}}& 
 -i\frac{g_{1}\mbox{sin}(t\sqrt{g_{1}^{2}+g_{2}^{2}})}
  {\sqrt{g_{1}^{2}+g_{2}^{2}}}& 
  \frac{g_{1}g_{2}\mbox{cos}(t\sqrt{g_{1}^{2}+g_{2}^{2}})-g_{1}g_{2}}
  {g_{1}^{2}+g_{2}^{2}}  \\
 -i\frac{g_{1}\mbox{sin}(t\sqrt{g_{1}^{2}+g_{2}^{2}})}
  {\sqrt{g_{1}^{2}+g_{2}^{2}}}& \mbox{cos}(t\sqrt{g_{1}^{2}+g_{2}^{2}})&
 -i\frac{g_{2}\mbox{sin}(t\sqrt{g_{1}^{2}+g_{2}^{2}})}
  {\sqrt{g_{1}^{2}+g_{2}^{2}}} \\
  \frac{g_{1}g_{2}\mbox{cos}(t\sqrt{g_{1}^{2}+g_{2}^{2}})-g_{1}g_{2}}
  {g_{1}^{2}+g_{2}^{2}}&
  -i\frac{g_{2}\mbox{sin}(t\sqrt{g_{1}^{2}+g_{2}^{2}})}
  {\sqrt{g_{1}^{2}+g_{2}^{2}}}&
  \frac{g_{2}^{2}\mbox{cos}(t\sqrt{g_{1}^{2}+g_{2}^{2}})+g_{1}^{2}}
  {g_{1}^{2}+g_{2}^{2}} 
  \end{array}
\right).
\end{eqnarray}
See $\S 3$ in \cite{FHKW2}.

\begin{Large}
\par \vspace{5mm} \noindent
{\bf Derivation of (\ref{eq:eigenvectors of Q})}
\end{Large}

\par \noindent
For $Q$ in (\ref{eq:three-level-matrix Q}) we consider the equations
\[
Q\mathbf{x}_{j}=\lambda_{j}\mathbf{x}_{j} \quad \mbox{for}\quad 1\leq j\leq 3.
\]
Namely,
\begin{equation}
\left(
  \begin{array}{ccc}
      0   & g_{1} & g_{3} \\
    g_{1} &   0   & g_{2} \\
    g_{3} & g_{2} &  0
  \end{array}
\right)
\left(
  \begin{array}{c}
   x \\
   y \\
   z
  \end{array}
\right)
=
\lambda_{j}
\left(
  \begin{array}{c}
   x \\
   y \\
   z
  \end{array}
\right)
\ \Longleftrightarrow\ 
\left\{
 \begin{array}{lll}
  g_{1}y+g_{3}z=\lambda_{j}x \\
  g_{1}x+g_{2}z=\lambda_{j}y \\
  g_{3}x+g_{2}y=\lambda_{j}z 
 \end{array}
\right.
\end{equation}

\par \noindent
We solve the above equations with respect to $z$. The result is 
\[
x=\frac{\lambda_{j}g_{3}+g_{1}g_{2}}{\lambda_{j}^{2}-g_{1}^{2}}z,\quad 
y=\frac{\lambda_{j}g_{2}+g_{1}g_{3}}{\lambda_{j}^{2}-g_{1}^{2}}z,
\]
where we have used the characteristic equation
\[
\lambda_{j}^{3}-(g_{1}^{2}+g_{2}^{2}+g_{3}^{2})\lambda_{j}-2g_{1}g_{2}g_{3}
=0,
\]
so 
\begin{equation}
\label{eq:A-0}
\mathbf{x}_{j}=z
\left(
 \begin{array}{c}
   \frac{\lambda_{j}g_{3}+g_{1}g_{2}}{\lambda_{j}^{2}-g_{1}^{2}} \\
   \frac{\lambda_{j}g_{2}+g_{1}g_{3}}{\lambda_{j}^{2}-g_{1}^{2}} \\
   1
 \end{array}
\right).
\end{equation}

\par \noindent
Next let us determine the normalization. 
\begin{equation}
\label{eq:A-1}
\langle{\mathbf{x}_{j}\vert\mathbf{x}_{j}}\rangle=1\ \Longleftrightarrow\ 
1=x^{2}+y^{2}+z^{2}
=
z^{2}
\frac{(\lambda_{j}g_{3}+g_{1}g_{2})^{2}+
(\lambda_{j}g_{2}+g_{1}g_{3})^{2}+(\lambda_{j}^{2}-g_{1}^{2})^{2}}
{(\lambda_{j}^{2}-g_{1}^{2})^{2}}
\end{equation}
From this 
\begin{eqnarray}
\label{eq:A-2}
&&(\lambda_{j}^{2}-g_{1}^{2})^{2}+(\lambda_{j}g_{3}+g_{1}g_{2})^{2}+
(\lambda_{j}g_{2}+g_{1}g_{3})^{2} \nonumber \\
=
&&\lambda_{j}^{4}+(-2g_{1}^{2}+g_{2}^{2}+g_{3}^{2})\lambda_{j}^{2}+
4g_{1}g_{2}g_{3}\lambda_{j}+g_{1}^{2}(g_{1}^{2}+g_{2}^{2}+g_{3}^{2}) 
\nonumber \\
=
&&\lambda_{j}^{4}-3g_{1}^{2}\lambda_{j}^{2}+
(g_{1}^{2}+g_{2}^{2}+g_{3}^{2})\lambda_{j}^{2}+
\underline{4g_{1}g_{2}g_{3}}\lambda_{j}+
g_{1}^{2}(g_{1}^{2}+g_{2}^{2}+g_{3}^{2}).
\end{eqnarray}
Now we want to remove the term $4g_{1}g_{2}g_{3}\lambda_{j}$. By using 
the characteristic equation 
$2g_{1}g_{2}g_{3}=\lambda_{j}^{3}-(g_{1}^{2}+g_{2}^{2}+g_{3}^{2})\lambda_{j}$, 
we have
\begin{eqnarray}
\label{eq:A-3}
\mbox{The right hand side of (\ref{eq:A-2})}
&=&3\lambda_{j}^{4}-3g_{1}^{2}\lambda_{j}^{2}-
(g_{1}^{2}+g_{2}^{2}+g_{3}^{2})\lambda_{j}^{2}+
g_{1}^{2}(g_{1}^{2}+g_{2}^{2}+g_{3}^{2}) \nonumber \\
&=&3\lambda_{j}^{2}(\lambda_{j}^{2}-g_{1}^{2})-
(g_{1}^{2}+g_{2}^{2}+g_{3}^{2})(\lambda_{j}^{2}-g_{1}^{2}) \nonumber \\
&=&(\lambda_{j}^{2}-g_{1}^{2})
(3\lambda_{j}^{2}-g_{1}^{2}-g_{2}^{2}-g_{3}^{2}).
\end{eqnarray}
From (\ref{eq:A-1})
\[
z=\frac{\sqrt{\lambda_{j}^{2}-g_{1}^{2}}}
{\sqrt{3\lambda_{j}^{2}-g_{1}^{2}-g_{2}^{2}-g_{3}^{2}}},
\]
so we have 
\begin{equation}
\label{eq:A-4}
\mathbf{x}_{j}=z
\left(
 \begin{array}{c}
   \frac{\lambda_{j}g_{3}+g_{1}g_{2}}{\lambda_{j}^{2}-g_{1}^{2}} \\
   \frac{\lambda_{j}g_{2}+g_{1}g_{3}}{\lambda_{j}^{2}-g_{1}^{2}} \\
   1
 \end{array}
\right)
=
\frac{1}{\sqrt{3\lambda_{j}^{2}-g_{1}^{2}-g_{2}^{2}-g_{3}^{2}}}
\left(
 \begin{array}{c}
   \frac{\lambda_{j}g_{3}+g_{1}g_{2}}{\sqrt{\lambda_{j}^{2}-g_{1}^{2}}} \\
   \frac{\lambda_{j}g_{2}+g_{1}g_{3}}{\sqrt{\lambda_{j}^{2}-g_{1}^{2}}} \\
   \sqrt{\lambda_{j}^{2}-g_{1}^{2}}
 \end{array}
\right)
\end{equation}
from (\ref{eq:A-0}). At this stage it is easy to conjecture 
\[
\frac{\lambda_{j}g_{3}+g_{1}g_{2}}{\sqrt{\lambda_{j}^{2}-g_{1}^{2}}}
=
\sqrt{\lambda_{j}^{2}-g_{2}^{2}},\quad 
\frac{\lambda_{j}g_{2}+g_{1}g_{3}}{\sqrt{\lambda_{j}^{2}-g_{1}^{2}}}
=
\sqrt{\lambda_{j}^{2}-g_{3}^{2}}.
\]
In fact,
\begin{eqnarray*}
&&\frac{\lambda_{j}g_{3}+g_{1}g_{2}}{\sqrt{\lambda_{j}^{2}-g_{1}^{2}}} \\
=&&
\sqrt{\frac{(\lambda_{j}g_{3}+g_{1}g_{2})^{2}}{\lambda_{j}^{2}-g_{1}^{2}}}
=
\sqrt{\frac{g_{3}^{2}\lambda_{j}^{2}+2g_{1}g_{2}g_{3}\lambda_{j}+
g_{1}^{2}g_{2}^{2}}{\lambda_{j}^{2}-g_{1}^{2}}}
=
\sqrt{\frac{(\lambda_{j}^{2}-g_{1}^{2})(\lambda_{j}^{2}-g_{2}^{2})}
{\lambda_{j}^{2}-g_{1}^{2}}}
=
\sqrt{\lambda_{j}^{2}-g_{2}^{2}},
\end{eqnarray*}
where we have used the equation
\begin{eqnarray*}
g_{3}^{2}\lambda_{j}^{2}+\underline{2g_{1}g_{2}g_{3}}\lambda_{j}+
g_{1}^{2}g_{2}^{2}
&=&
g_{3}^{2}\lambda_{j}^{2}+
\lambda_{j}^{4}-(g_{1}^{2}+g_{2}^{2}+g_{3}^{2})\lambda_{j}^{2}+
g_{1}^{2}g_{2}^{2}  \\
&=&
\lambda_{j}^{4}-(g_{1}^{2}+g_{2}^{2})\lambda_{j}^{2}+g_{1}^{2}g_{2}^{2} \\
&=&(\lambda_{j}^{2}-g_{1}^{2})(\lambda_{j}^{2}-g_{2}^{2}).
\end{eqnarray*}
Similarly, we obtain
\[
\frac{\lambda_{j}g_{2}+g_{1}g_{3}}{\sqrt{\lambda_{j}^{2}-g_{1}^{2}}}
=
\sqrt{\lambda_{j}^{2}-g_{3}^{2}}.
\]

\par \noindent
From (\ref{eq:A-4}) we finally obtain
\[
\mathbf{x}_{j}=
\frac{1}{\sqrt{3\lambda_{j}^{2}-g_{1}^{2}-g_{2}^{2}-g_{3}^{2}}}
\left(
 \begin{array}{c}
   \sqrt{\lambda_{j}^{2}-g_{2}^{2}} \\
   \sqrt{\lambda_{j}^{2}-g_{3}^{2}} \\
   \sqrt{\lambda_{j}^{2}-g_{1}^{2}}
 \end{array}
\right)
=
\left(
 \begin{array}{c}
  \sqrt{\frac{\lambda_{j}^{2}-g_{2}^{2}}{3\lambda_{j}^{2}-g_{1}^{2}-g_{2}^{2}-
  g_{3}^{2}}} \\
  \sqrt{\frac{\lambda_{j}^{2}-g_{3}^{2}}{3\lambda_{j}^{2}-g_{1}^{2}-g_{2}^{2}-
  g_{3}^{2}}} \\
  \sqrt{\frac{\lambda_{j}^{2}-g_{1}^{2}}{3\lambda_{j}^{2}-g_{1}^{2}-g_{2}^{2}-
  g_{3}^{2}}}
 \end{array}
\right)\equiv \ket{\lambda_{j}}.
\]

A comment is in order. $\sqrt{t}$ means the complex square root defined by\ 
$
\sqrt{t}=\mbox{e}^{i\arg{t}/2}\sqrt{|t|}
$
, so in our case we must choose its sign (for example, $\sqrt{4}=2\ \mbox{or} 
-2$) carefully.

\vspace{10mm}

\end{document}